\documentclass[fleqn,a4paper,11pt]{article}
 
\usepackage{color}
 
\usepackage{graphicx}
\usepackage{amssymb}
\newcommand \pt {p$_T$}

\newcommand \sqn {$\sqrt{s_{_{NN}}}$ }
 
\newcommand \dau {d+Au }
\newcommand \auau {Au+Au }
\newcommand \pbpb {Pb+Pb }
\newcommand \cucu {Cu+Cu }
\newcommand \sqnr {$\sqrt{s_{_{NN}}}$~=~200~GeV}

\newcommand \raa {R$_{AA}$ }
\newcommand \vtwo {v$_2$}

\def\JPG{{J. Phys.}~{\bf G}}

\def\NPA{{Nucl. Phys.}~{\bf A}}

\def\PRL{Phys. Rev. Lett.\ }
\def\PR{Phys. Rev.\ }

\def\PRC{{Phys. Rev.}~{\bf C}}

\def\EPJC{{Eur.~Phys.~J.}~{\bf C}}

\begin{document}
\title{\bf {The Strongly Interacting Quark Gluon Plasma at RHIC and LHC}}
 
 \author{Itzhak Tserruya \footnote{email: Itzhak.Tserruya@weizmann.ac.il} \\   
                    Weizmann Institute of Science, Rehovot, Israel} 
 
\maketitle 
    
\begin{abstract}
 The study of heavy-ion collisions has currently unprecedented opportunities with two first class facilities,  the Relativistic Heavy Ion Collider (RHIC) at BNL and the Large Hadron Collider (LHC) at CERN, and five large experiments ALICE, ATLAS, CMS, PHENIX and STAR producing a wealth of high quality data.  Selected results recently obtained are presented  on the study of flow, energy loss and direct photons.
\end{abstract}

\section{Introduction}
\label{intro}
Since about two years, the field of relativistic heavy-ion collisions is in a unique and unprecedented situation. It benefits from a wealth of high quality results obtained by five large experiments in operation at two outstanding and complementary facilities. RHIC started operation in the year 2000 and has demonstrated unique flexibility and capabilities in delivering beams at a very broad energy range from the top energy at  \sqnr~down to \sqn = 7 GeV and providing collisions of a large variety of species pp, dA, CuCu, AuAu and recently U+U as well as asymmetric collisions like Cu+Au. The LHC at CERN started a most fruitful heavy ion program in 2010 opening a new energy frontier with the study of Pb+Pb collisions at \sqn =  2.76 TeV. The program  is carried out by five large experiments, PHENIX and STAR at RHIC and ALICE, ATLAS and CMS at the LHC. The emphasis is on precise studies to characterize the properties of the strongly interacting quark gluon plasma (sQGP) discovered at RHIC. The study of many observables over two to three orders of magnitude in energy provide strong contraints and valuable guidance in the development of theoretical models.  In the limited space of this paper it is not possible to do justice to all the recent developments in the field. The paper is therefore restricted to recent results on a few topics including flow, energy loss and direct photons.
 
\section{Flow}
\label{sec:flow}
The study of flow is a main focus of interest. It has the potential to address tranport properties of the plasma like the shear viscosity to entropy ratio ($\eta$/s), the speed of sound etc. Of particular interest is the detailed comparison of the flow parameters at RHIC and LHC energies which could reveal differences in the plasma formed at these two energy regimes.

\begin{figure}[h!]
\begin{center}
     \includegraphics[width=46mm, height=40mm]{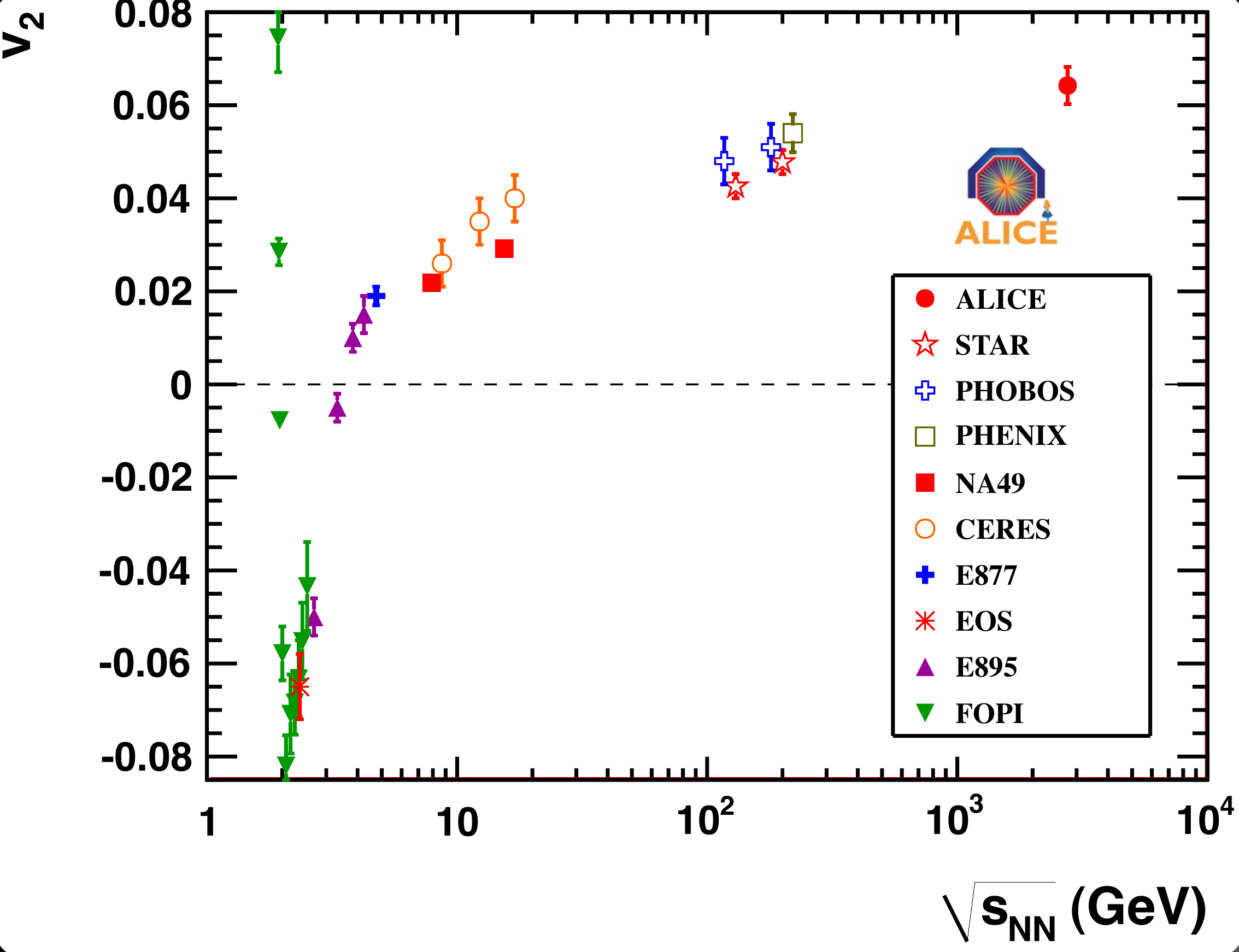}
     \includegraphics[width=46mm, height=40mm]{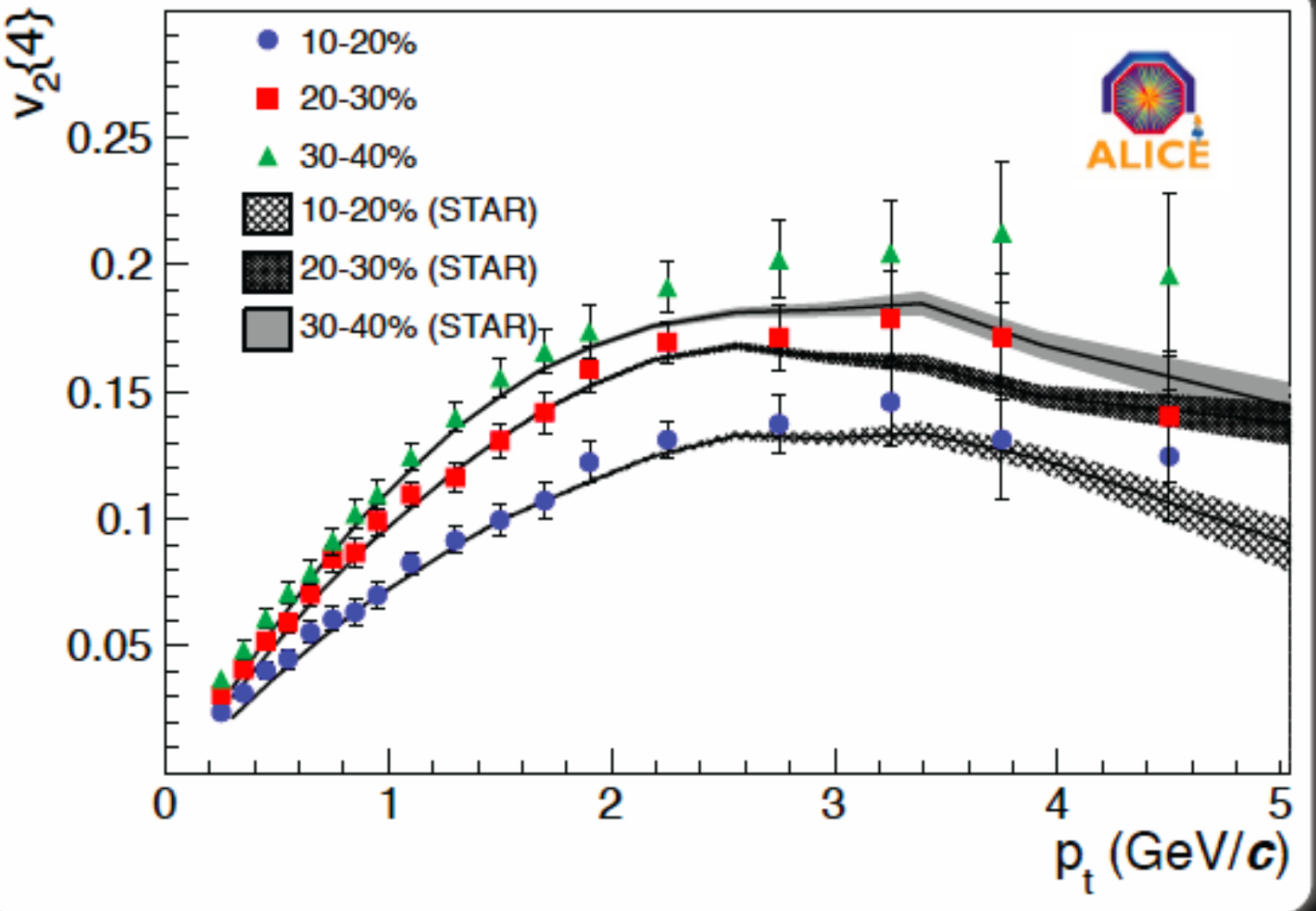}
     \includegraphics[width=46mm, height=40mm]{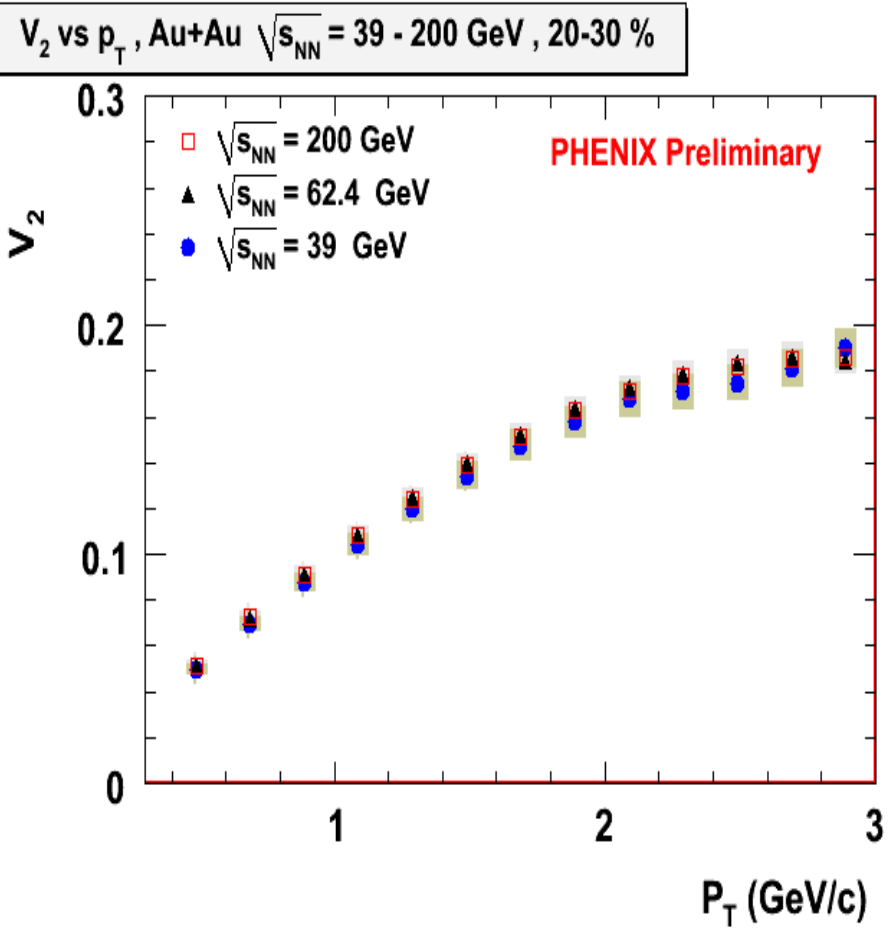} 
  \caption{Average elliptic flow \vtwo~vs. \sqn (left panel) \cite{alice-flow},  comparison of \vtwo(\pt) results for charged hadrons at RHIC (STAR) and LHC (ALICE) energies (middle panel)   \cite{alice-flow} and differential elliptic flow \vtwo(\pt) of charged hadrons in 20-30\% \auau collisions at \sqn = 39-200 GeV (right panel) \cite{gong-qm11} .}
  \label{fig:v2-roots}
\end{center}
\vspace{-4mm}
 \end{figure}

The average elliptic flow 
 of inclusive charged hadrons in the 20-30\% centrality class increases by  $\sim$25\% from RHIC to LHC (see left panel of Fig.~\ref{fig:v2-roots}) \cite{alice-flow}. This increase mainly reflects the observed increase in the average p$_T$ of the charged hadrons rather than an increase of the differential elliptic flow \vtwo(\pt). 
The latter is in fact remarkably constant from RHIC to LHC as demonstrated in the middle panel of Fig.~\ref{fig:v2-roots} that compares the differential elliptic flow of charged hadrons measured by ALICE and STAR \cite{alice-flow}. The differential elliptic flow of charged hadrons is also remarkably constant at  lower energies, from 200 down to 39 GeV, as demonstrated by the PHENIX measurements \cite{gong-qm11} shown in the right panel of Fig.~\ref{fig:v2-roots}. The saturation of \vtwo(\pt) at, or below, \sqn = 39 GeV suggests that the perfect liquid property of the sQGP discovered at RHIC is valid from this low energy up to at least 2.76~TeV.
At RHIC, the saturation of \vtwo(\pt) for charged hadrons  is also observed for identified hadrons ($\pi, K, p$) in the energy range \sqn = 39-200 GeV. At LHC the situation is less clear. The observed saturation of \vtwo(\pt)  for unidentified charged hadrons could be fortuitous and result from higher \vtwo(\pt) values at the LHC compared to RHIC for $\pi$ and and $K$ compensated by lower \vtwo(\pt) values for $p$ \cite{roy-2012}. The differences are relatively small, of the order of 20\%, but if confirmed they could signal small differences in the properties of the sQGP formed at the LHC compared to the one formed at RHIC.  

 The valence quark scaling observed at RHIC seems also to work at the LHC although deviations are observed for protons \cite{snellings-qm11}.
 
\begin{figure}[h!]
\begin{center}
    \includegraphics[width=55mm, height=75mm]{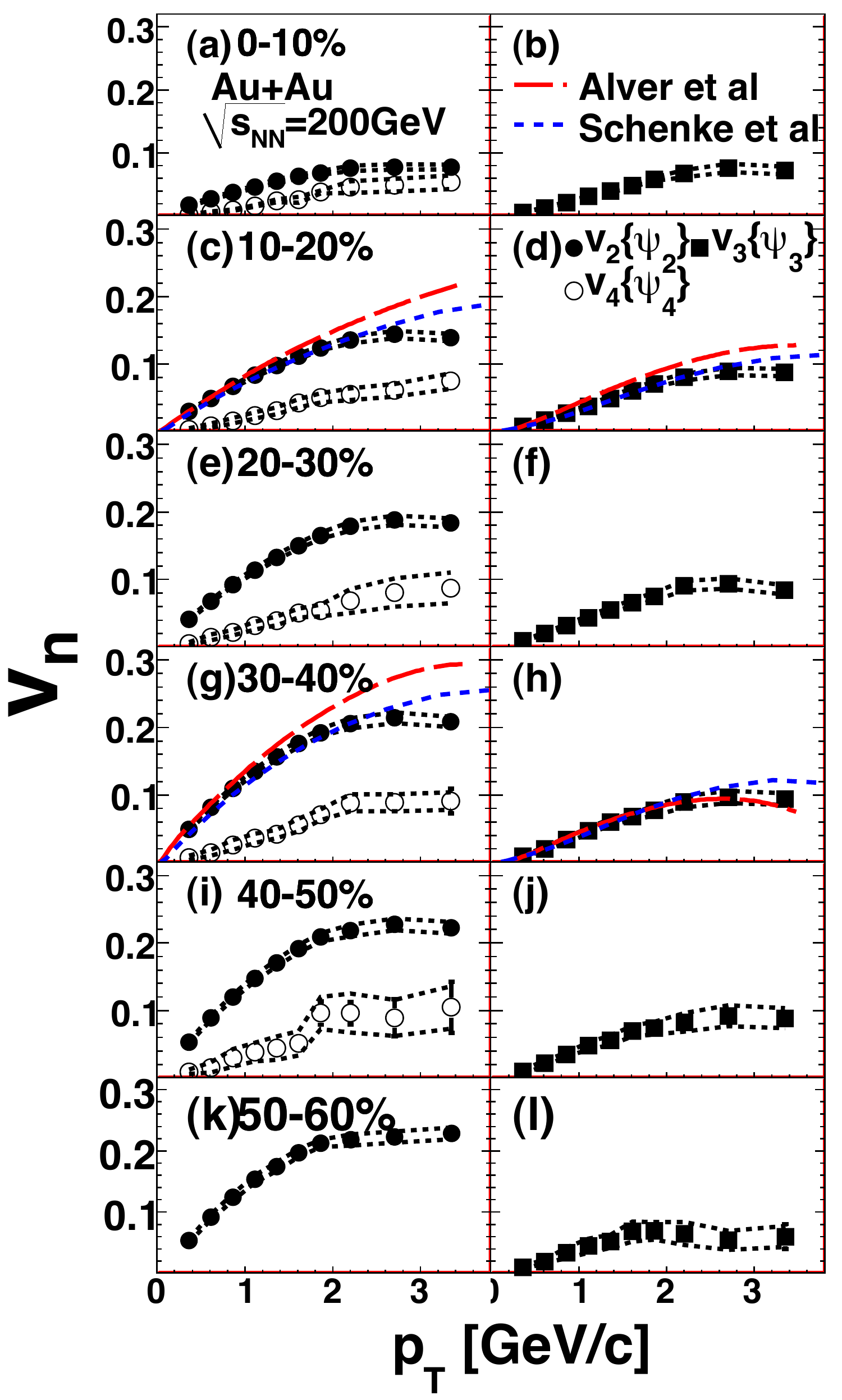}
\hspace{10mm}
     \includegraphics[width=55mm, height=75 mm]{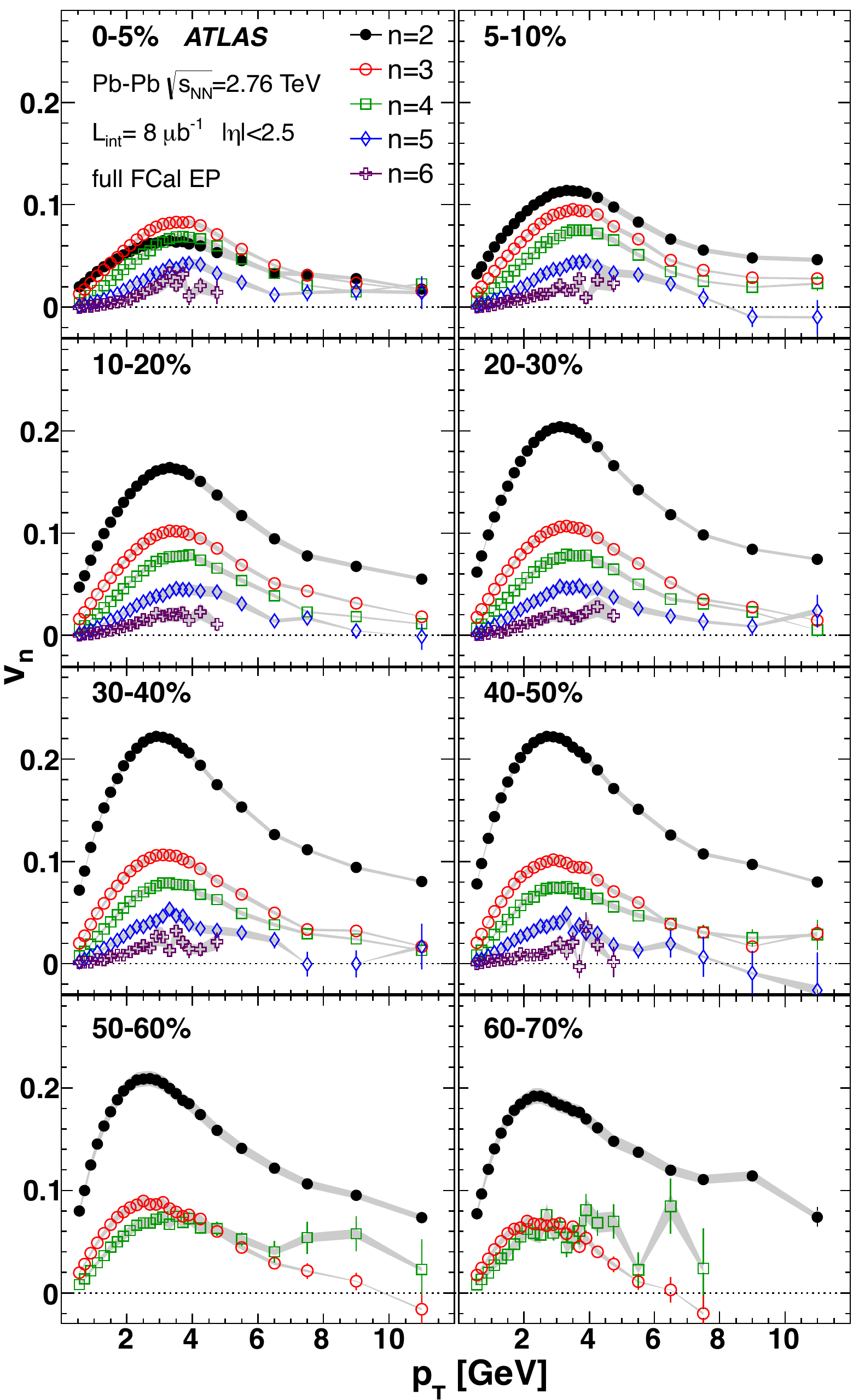}
    \caption{Higher order flow harmonics measured by PHENIX (left panel) \cite{vn-phenix} and ATLAS (right panel) \cite{vn-atlas} for different centrality classes in collisions of \auau at \sqnr~and \pbpb at \sqn = 2.76 TeV, respectively.}
  \label{fig:high-harmonics}
\end{center}
\vspace{-10mm}
 \end{figure}

Motivated by recent theoretical work \cite{flow-theory-high-vn},
 the large five experiments have performed over the past couple of years systematic measurements showing the importance of the higher order harmonic flow components.  
The characteristic features are very similar at RHIC and LHC (see Fig.~\ref{fig:high-harmonics}) \cite{vn-phenix, vn-atlas}:
\begin{itemize} 
\vspace{-2.0mm}
\item 
sizable v$_n$ up to the sixth order;
\vspace{-2.0mm}
\item 
same pattern for all n: v$_n$ rises up to $\sim$3~GeV/c and then falls at higher p$_T$;
\vspace{-2.0mm}
\item 
weaker or no centrality dependence of v$_3$-v$_6$ in contrast to v$_2$ that exhibits a strong centrality dependence.
\end{itemize}
The saturation of v$_2$  mentioned above also holds for the higher flow components. 

The importance of the higher order harmonics has crucial consequences. The long range $\Delta \eta$ near-side correlations (the so-called ridge) and the double peak structure in the away-side correlations (known as the double-hump), both observed in two-particle correlations, largely disappear once the higher order flow components are subtracted \cite{vn-atlas}. These low-\pt~correlation effects reflect fluctuations in the initial geometry rather than the response of the medium to the energy deposited by high energy partons.  The higher order harmonics and in particular v$_3$ provide additional constraints to the models and should help determining the shear viscosity to entropy ratio $\eta$/s with higher precision \cite{vn-phenix}.

\section{Energy loss}
\label{sec:energy-loss}
The jet quenching phenomena is a topic of great interest. Their study greatly benefits from the higher energy reach offered by the LHC. By jet quenching one refers to the energy loss of hard scattered partons as they traverse the color medium formed in ultrarelativistic heavy ion collisions. The goal is to accurately measure this energy loss. This is not a simple task as it requires knowledge of the initial parton energy, the path length in the medium and the final parton energy. A variety of observables are used, the suppression of high \pt~particles, two-particle correlations and full jet reconstruction. Some of the most recent results are reviewed below.

\subsection{Suppression of high \pt~ particles - \raa}
\label{sec:raa}
\begin{figure}[h!]
     \includegraphics[width=72mm, height=55mm]{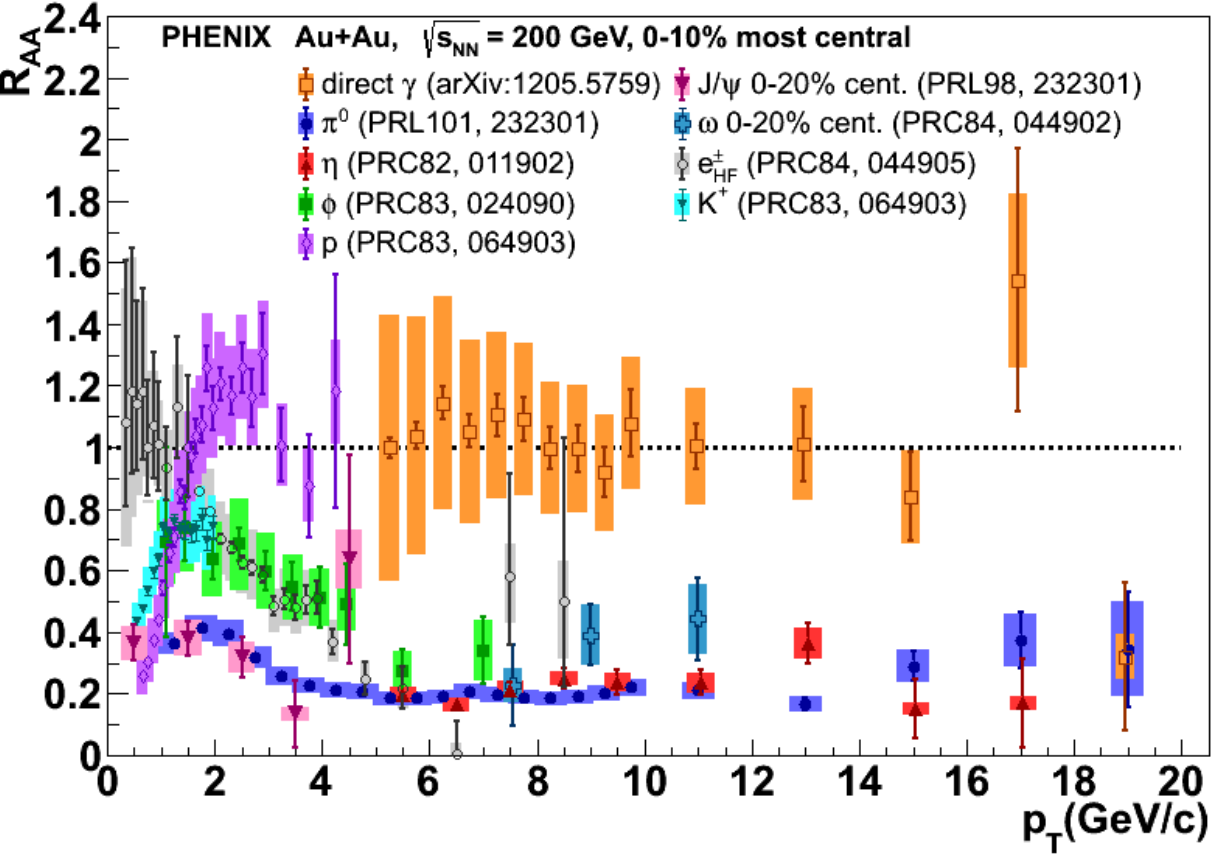}
     \includegraphics[width=72mm, height=55mm]{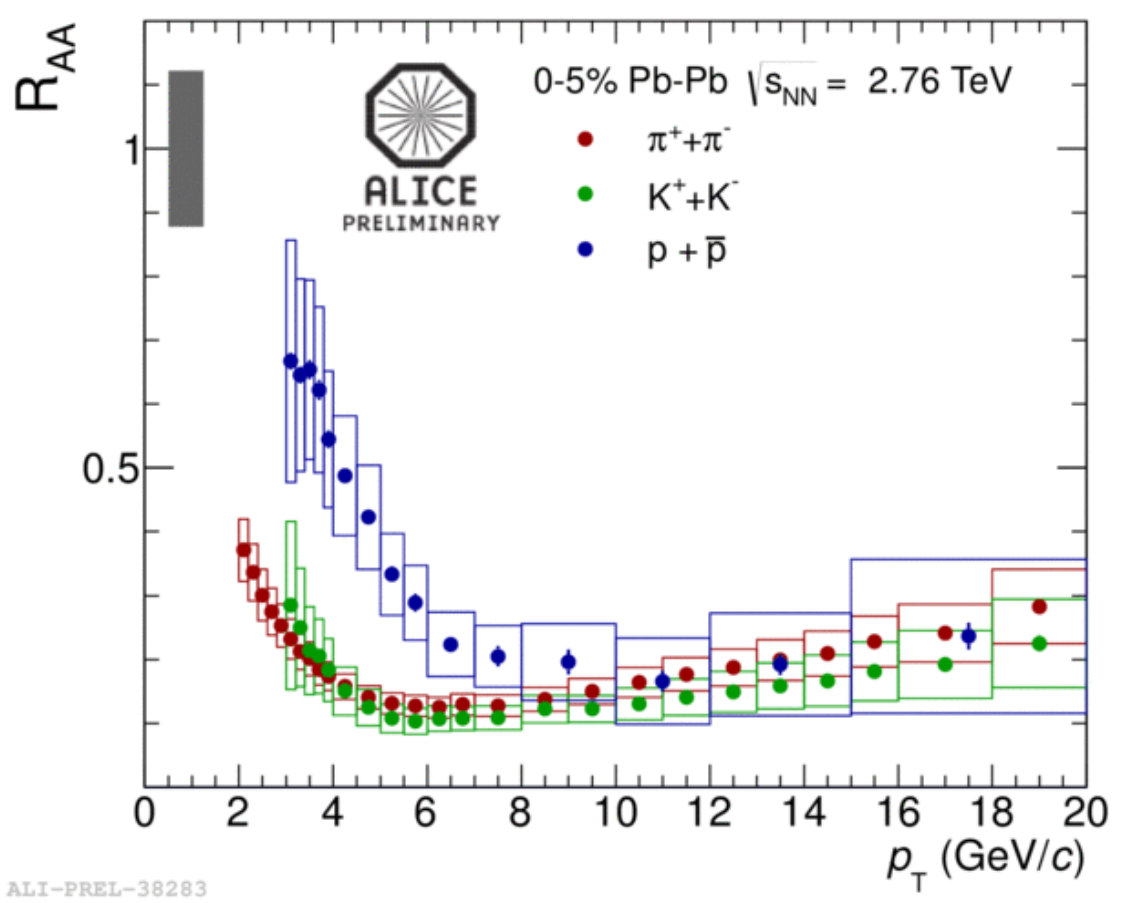}
     \vspace{-3mm}
    \caption{Left panel: \raa for baryons, strange mesons, e from heavy flavor,  light quark mesons and direct photons measured by PHENIX in 0-10\% most central Au+Au collisions at \sqn = 200 GeV  \cite{raa-zoo-phenix}. Right panel: \raa of $\pi$, K and p measured by ALICE in central Pb+Pb collisions at \sqn = 2.76 TeV \cite{raa-alice}.}
  \label{fig:raa-all-particles}
\vspace{-2mm}
 \end{figure}

 The discovery of jet quenching phenomena was first made at RHIC in measurements of the suppression of high \pt~particles \cite{phenix-suppression}. 
The  suppression is quantified by the nuclear modification factor, \raa, defined as the ratio of the particle yield in nucleus-nucleus collisions, scaled down by the number of binary collisions, to the particle yield in pp collisions. Over the past decade the RHIC experiments  have systematically measured the suppression of a large variety of particles that challenge and constrain energy loss models. The left panel of Fig.~\ref{fig:raa-all-particles} shows an updated compilation of \raa measurements  by the PHENIX experiment \cite{raa-zoo-phenix}. At low \pt~ (\pt~$<$ 6-7 GeV/c), the results show an interesting hierarchy in the suppression pattern of identified particles.  Direct photons, as expected, are not suppressed.  Light quark mesons show the largest suppression whereas baryons have very small or no suppression at all.  Strange mesons and electrons from heavy flavor ($e_{HF}$) show intermediate suppression:   
\begin{center}
\vspace{-2mm} 
{\raa (light quark mesons) \(<\) \raa (strange mesons and $e^\pm$ from HF) \(<\) \raa (baryons) }
\vspace{-2mm} 
\end{center}
On the other hand, at higher p$_T$, all particles, baryons and mesons, independently of their quark flavor, are strongly suppressed and seem to exhibit the same suppression level. 

Systematic measurements of identified particles \raa are also beeing performed at the LHC. The ALICE results available so far, including \raa of $\pi$, K and p  (see right panel of Fig.~\ref{fig:raa-all-particles}) \cite{raa-alice} and \raa of D mesons \cite{raa-alice-dmesons} seem to exhibit a similar pattern. 

\begin{figure}[h!]
     \includegraphics[width=72mm, height=55mm]{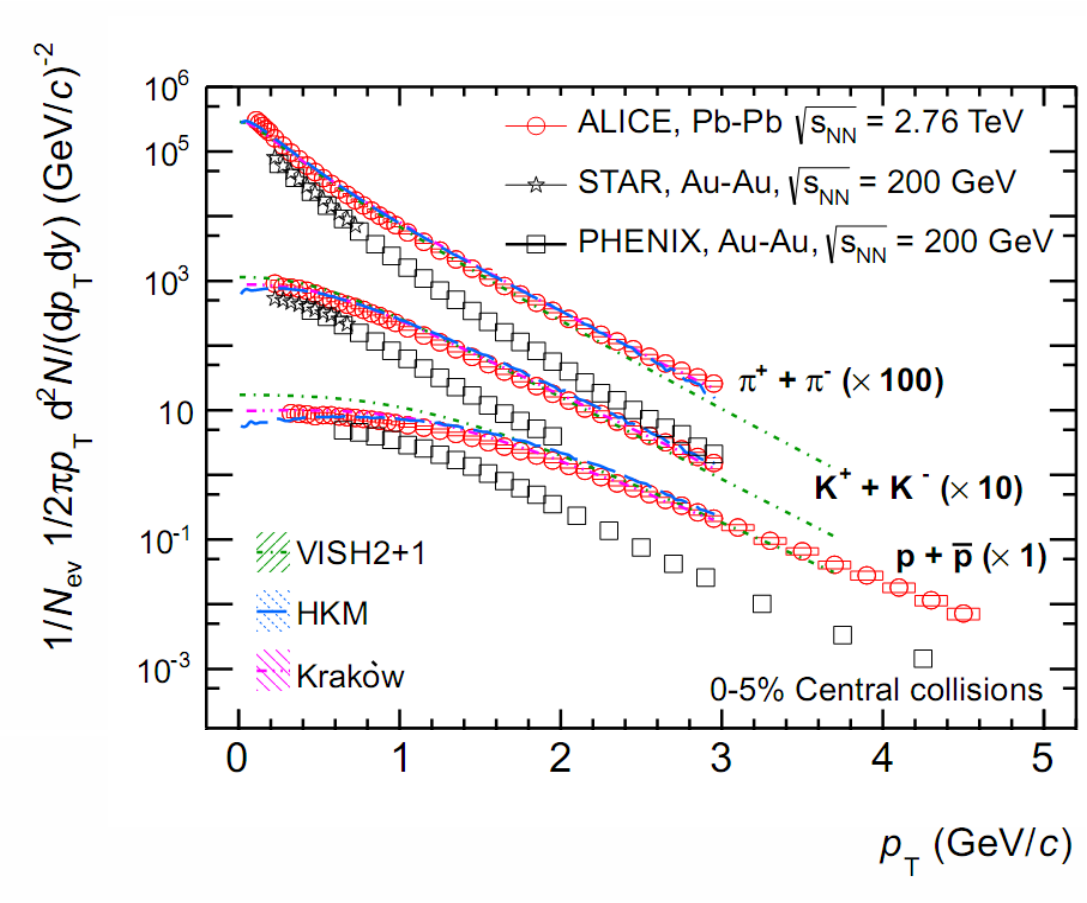}
     \includegraphics[width=72mm, height=55mm]{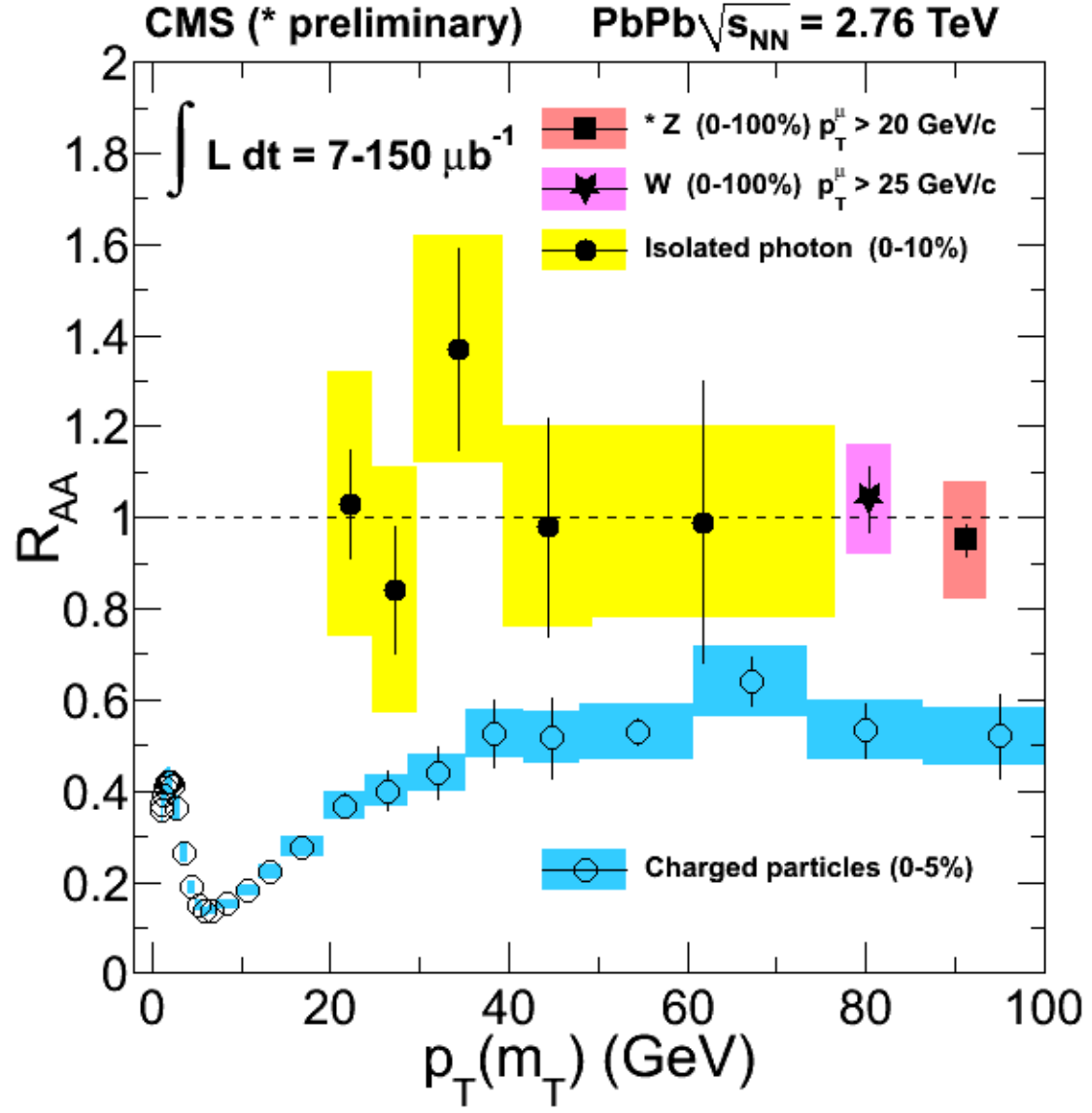}
     \vspace{-3mm}
\caption{Left panel: Identified particle spectra in central collisions at \sqn = 200 GeV from PHENIX and STAR and at \sqn = 2.76 TeV from ALICE \cite{alice-spectra}. Right panel: \raa of charged particles from CMS in central \pbpb collisions at \sqn = 2.76 TeV \cite{raa-hadrons-cms}.}
  \label{fig:raa-cms}
 \vspace{-6mm}
 \end{figure}

The LHC brings a new dimension to the study of R$_{AA}$. Particle spectra can be measured with unprecedented \pt~reach, up to 100~GeV/c. This is possible since the particle production cross sections are considerably higher at the LHC,  and the particle spectra are harder at the LHC than at RHIC reflecting a higher radial flow (see left panel of Fig.~\ref{fig:raa-cms}) \cite{alice-spectra}. 
The \raa of charged particles measured by the CMS in central \pbpb collisions at \sqn = 2.76 TeV  is displayed in the right panel of Fig.~\ref{fig:raa-cms}.  
At low p$_T$,  \raa exhibits first a fast rise followed by a gradual decrease, reaching  maximum suppression at p$_T$~=~6-7~GeV/c. At \pt~$>$ 7 GeV/c, \raa steadily increases up to $\sim$30 GeV/c  and appears to saturate at higher \pt~values.
 In the \pt~region of overlap, RHIC data (see e.g. the $\pi^0$ data in the left panel of Fig.~\ref{fig:raa-all-particles}) show a similar pattern, although the minimum at \pt~=~6-7~GeV/c is less deep and the rise at \pt~$>$ 7 GeV/c is not clearly established beyond the experimental uncertainties.

\subsection{Jets and correlations}
\label{sec:jets}
The study of jets is a prominent part of the heavy ion experiments at RHIC and LHC. Jets provide a powerful tool for directly studying the energy loss of high energy partons in the hot and dense medium formed in heavy ion collisions. The energy lost could manifest itself in the properties of the jet by e.g. changes in the jet shape and the jet fragmentation function.

There are obvious differences between  jets at RHIC and LHC. At RHIC, jet studies are practically limited to energies below $E_{jet}$~=~30-50~GeV, the limit being imposed by the jet production cross section. Furthermore, jet information is mainly derived via particle correlation studies. It is only recently that both PHENIX and STAR were able to perform full jet reconstruction. 
At LHC, on the other hand, jet studies are carried out for high energy jets (E$_{jet}$~\(>\)~25~GeV) which are copiously produced, prominently visible and can be fully reconstructed, whereas low energy jets cannot be reconstructed due to fluctuations in the underlying event. An interesting question is whether the quench phenomena of low energy jets observed at RHIC are qualitatively different from those of  high energy jets measured at LHC. In the following, the main characterictic features of jets at RHIC and LHC are compared.

\begin{figure}[h!]
     \includegraphics[width=72mm, height=55mm]{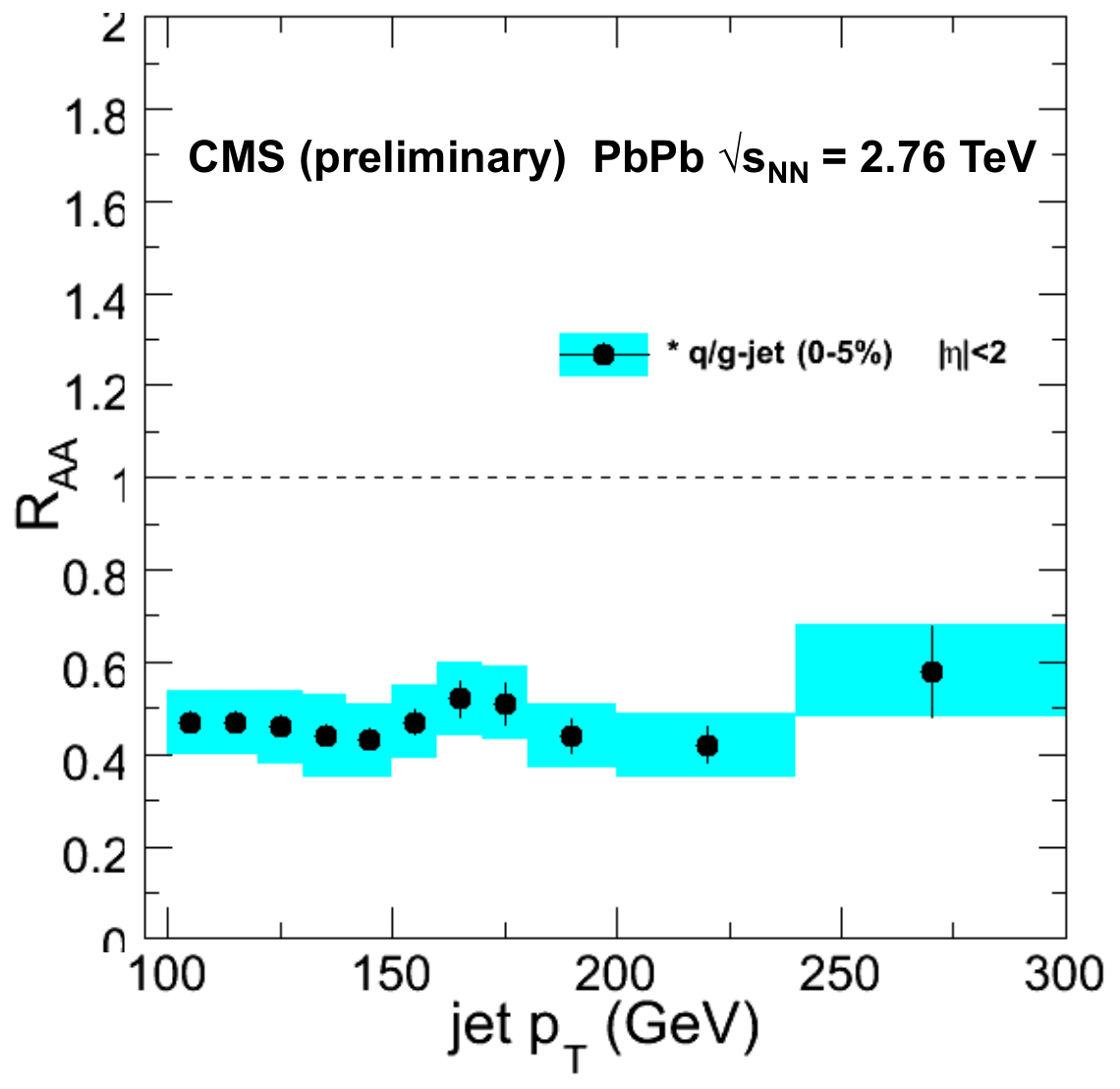}
     \includegraphics[width=72mm]{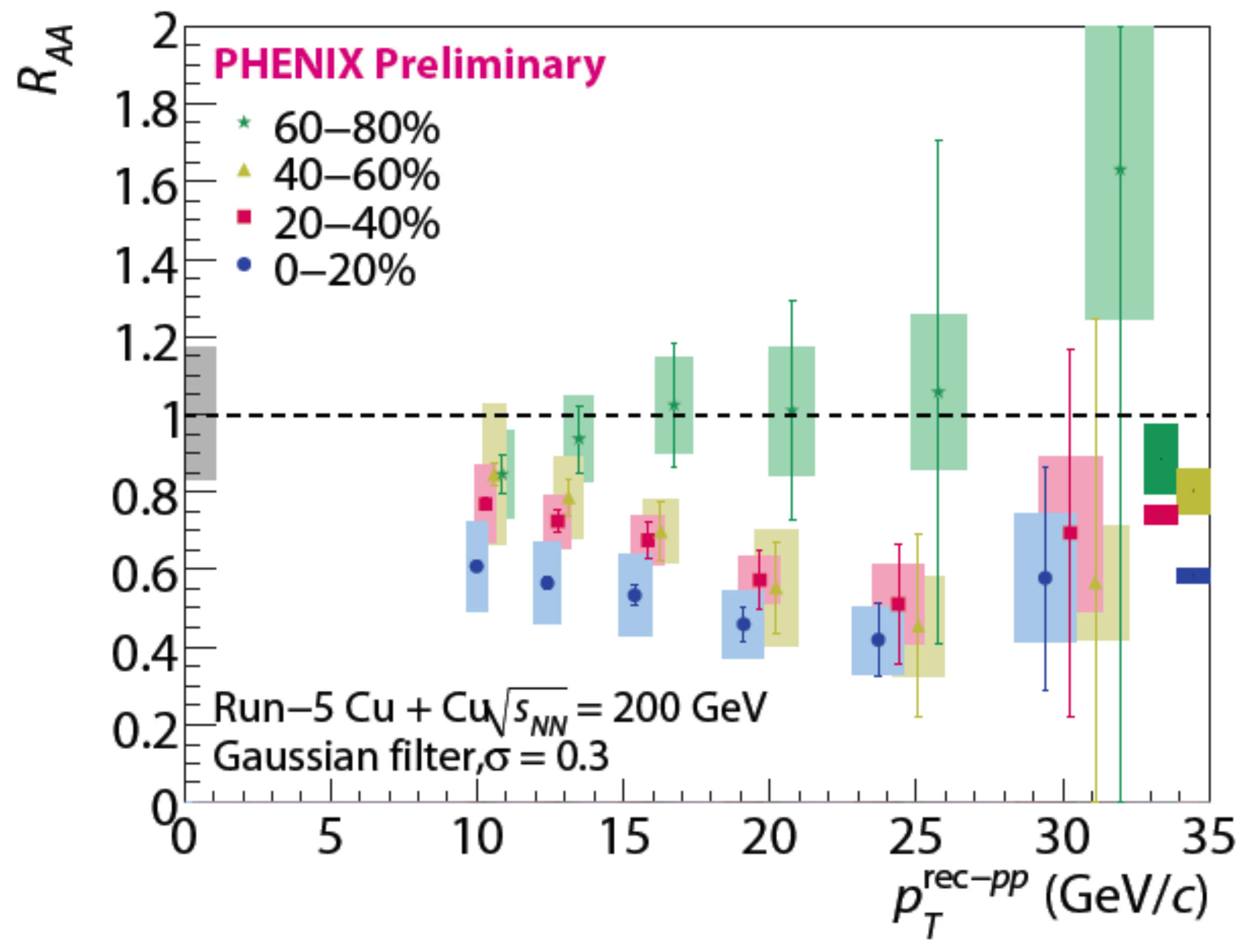}
    \vspace{-3mm}
            \caption{Jet suppression in central \pbpb collisions at \sqn = 2.76 TeV from CMS (left panel) \cite{jets-raa-cms} and in several centrality ranges in \cucu collisions at \sqn = 200 GeV from PHENIX \cite{jets-raa-phenix}. }
     \label{fig:jets-raa}
 \end{figure}

\vspace{3mm}
$\bullet$ \underline {Jet suppression.}
Jets appear to be equally suppressed at RHIC and LHC. 
At LHC, the jet yield is suppressed in central collisions by about a factor of 2 and the suppression level is independent of jet \pt~from 100 up to 300 GeV/c (see left panel of Fig.~\ref{fig:jets-raa} \cite{jets-raa-cms}).
A similar level of suppression is observed for low \pt~jets in central collisions at RHIC  as shown in the right panel of Fig.~\ref{fig:jets-raa} \cite{jets-raa-phenix}.  

\begin{figure}[h!]
     \includegraphics[width=60mm, height=40mm]{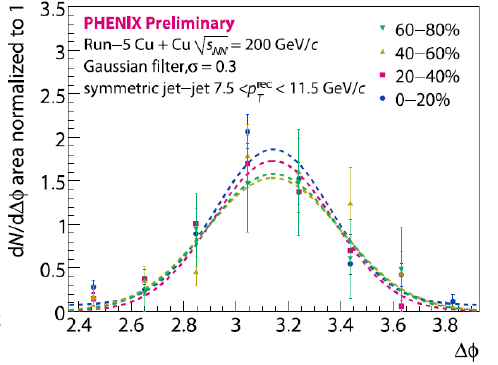}
     \includegraphics[width=80mm, height=40mm]{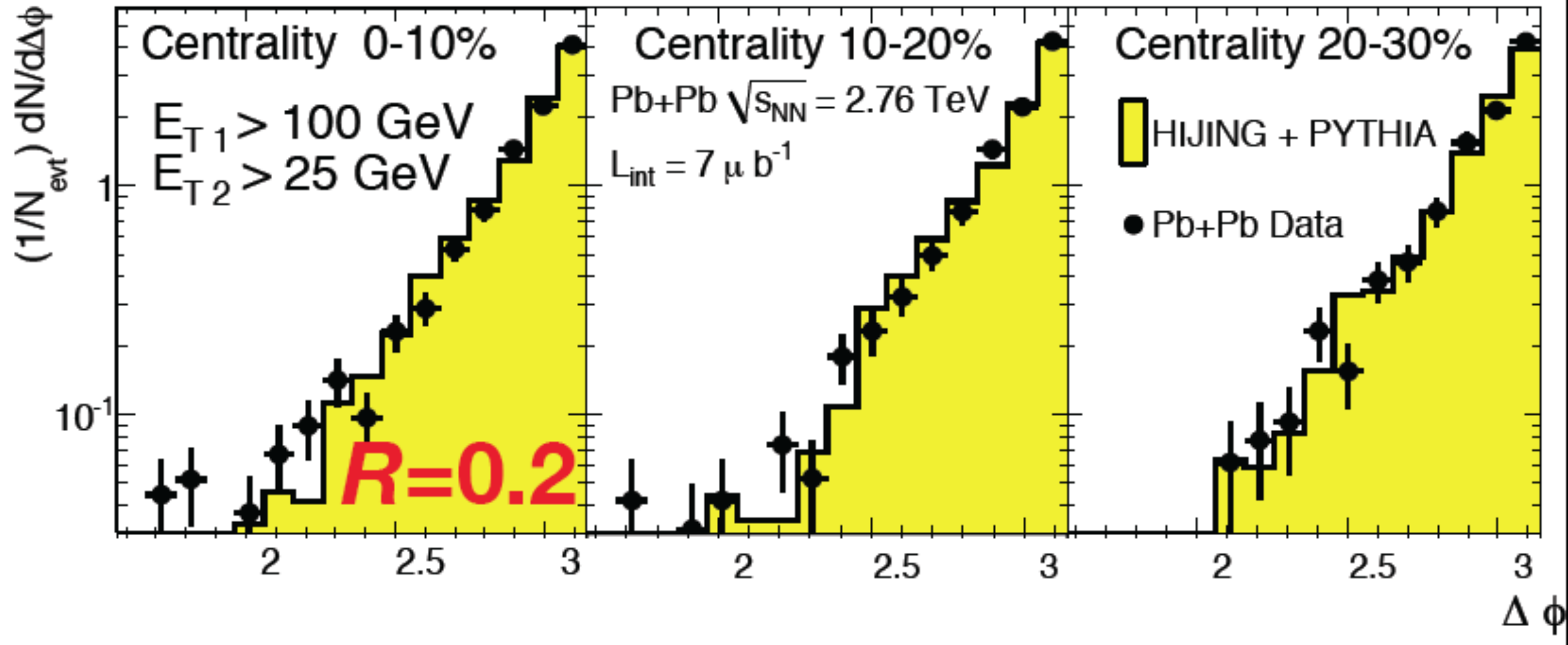}
     \vspace{-3mm}
            \caption{Angular correlations of dijets ($\Delta \phi = \phi_{jet1} - \phi_{jet2}$) in Cu+Cu collisions at RHIC from PHENIX (left panel) \cite{jets-back-to-back-phenix} and in Pb+Pb collisions at LHC from ATLAS (right panel) \cite{jets-atlas}.}
     \label{fig:jets-back-to-back}
     \vspace{-2mm}
 \end{figure}

$\bullet$ \underline {Dijet angular correlations.}
Dijets are mostly back-to-back in AA collisions both at the LHC and RHIC as shown in Fig.~\ref{fig:jets-back-to-back} \cite{jets-back-to-back-phenix, jets-atlas}. They show the same angular correlations as in pp  for all centralities and there is no evidence for deflection of the away-side jet.

\begin{figure}[h!]
\begin{center}
     \includegraphics[width=60mm, height=50mm]{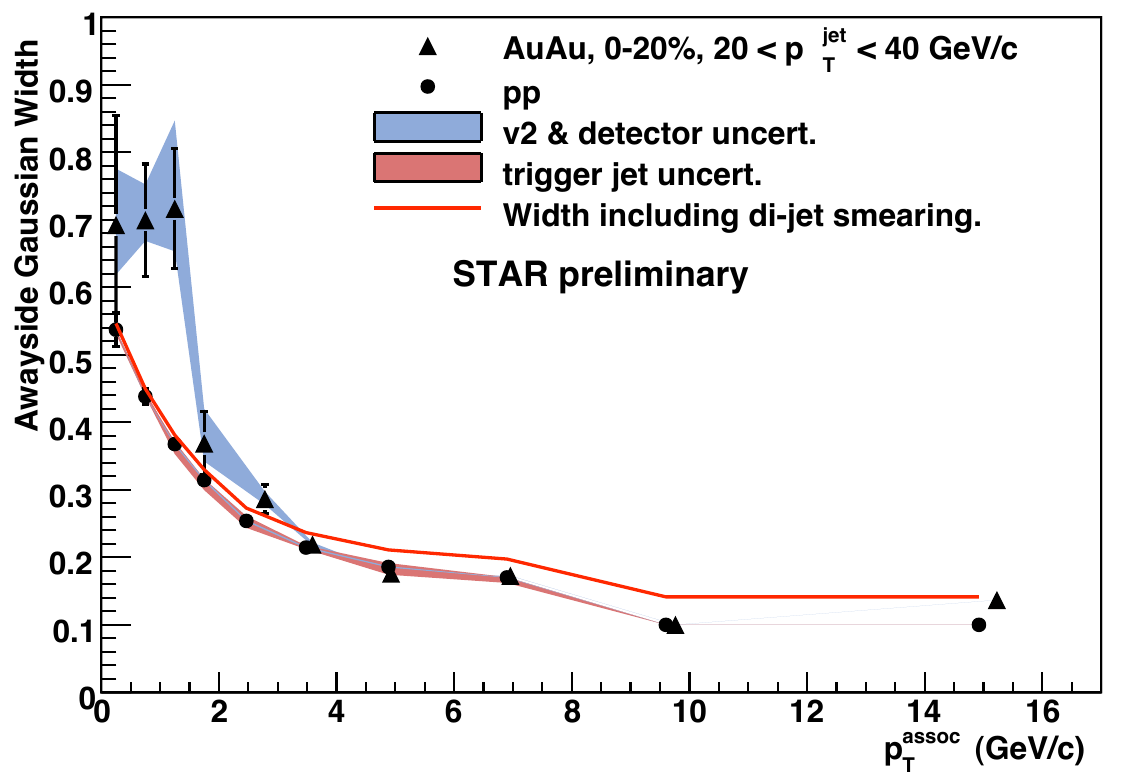}
\hspace{7mm}  
     \includegraphics[width=60mm, height=50mm]{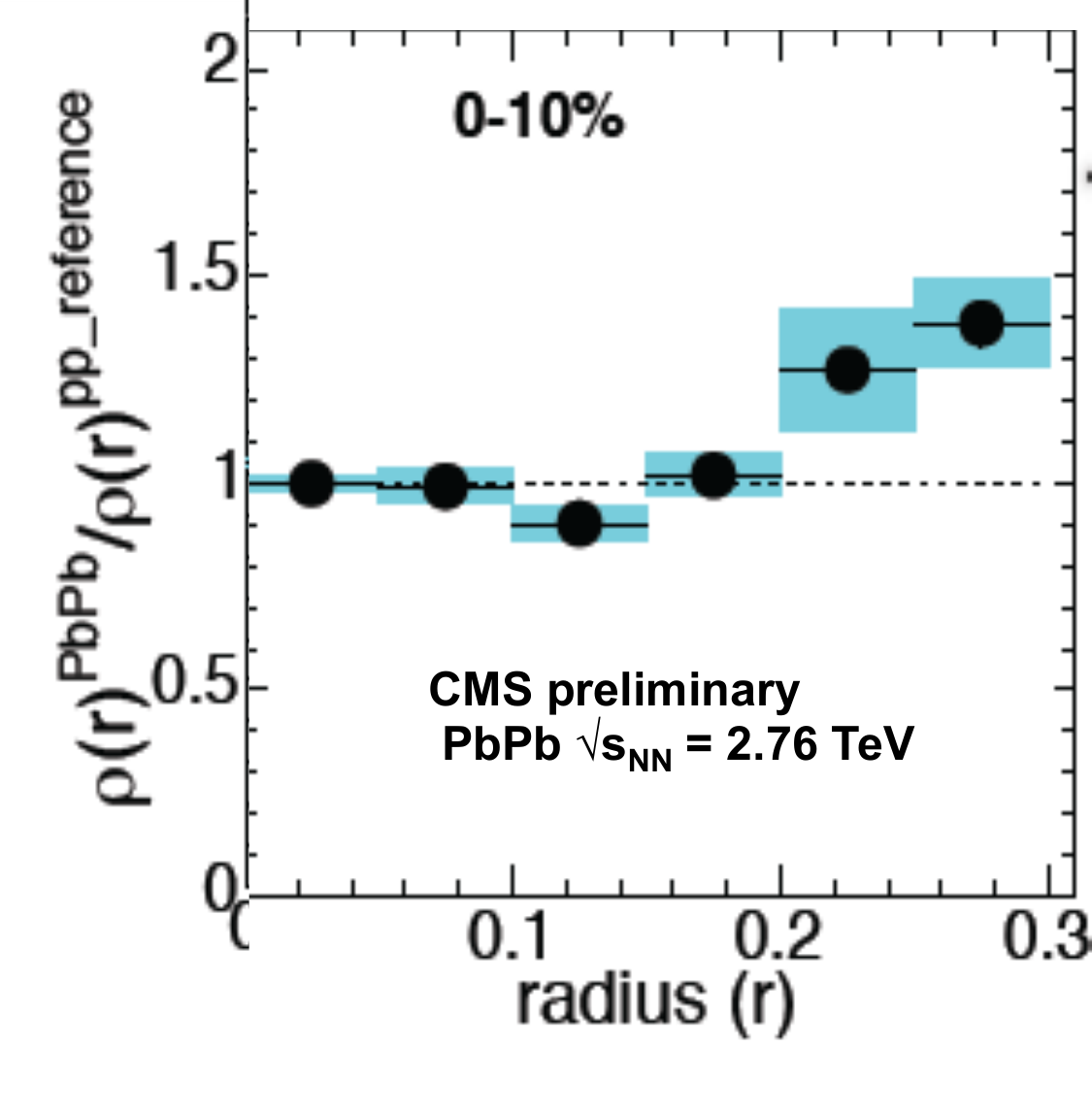}
\end{center}
\vspace{-6mm}
            \caption{Left panel: width of the away-side jet-hadron correlations in central Au+Au and pp collisions as function of \pt~of the associated hadrons from STAR \cite{caines-qm11}. Right panel: ratio of the differential jet shape in central \pbpb to pp collisions as function of the invariant radius r from the jet axis \cite{jet-shape-ff-cms}.}
     \label{fig:jet-broadening}
 \end{figure}

$\bullet$ \underline {Jet broadening.}
The jet shape is modified in nuclear collisions compared to pp collisions both at RHIC and LHC as illustrated in Fig.~\ref{fig:jet-broadening}.
The left panel shows results derived from jet-hadron correlations measured by STAR in central Au+Au collisions \cite{caines-qm11}. 
For high \pt~associated particles, the width of the away-side correlations is the same as in pp collisions, whereas it is considerably larger, by almost a factor of 2, when selecting low \pt~associated particles. This suggests that the energy lost by the parton as it traverses the medium appears in soft hadrons that broaden the recoiling jet shape but remain correlated with the original parton direction.
Similar  conclusions can be reached from the plot on the right panel that shows CMS results of the ratio of the differential jet shape in central \pbpb to pp collisions for inclusive jets with \pt~$>$ 100 GeV/c and charged particles with \pt~$>$ 1 GeV/c \cite{jet-shape-ff-cms}. The differential jet shape is defined as the fraction of the jet transverse momentum contained in an annulus of invariant radius r around the jet axis and is determined using charged tracks only.
A significant increase of $\sim$40\%  is observed at large values of r in the most central \pbpb collisions shown in the right panel indicating a broadening of the jet shape, whereas the ratio is consistent with unity for peripheral collisions. Additional measurements show  that this broadening is due to soft particles.

\begin{figure}[h!]
\begin{center}
     \includegraphics[width=95mm, height=50mm]{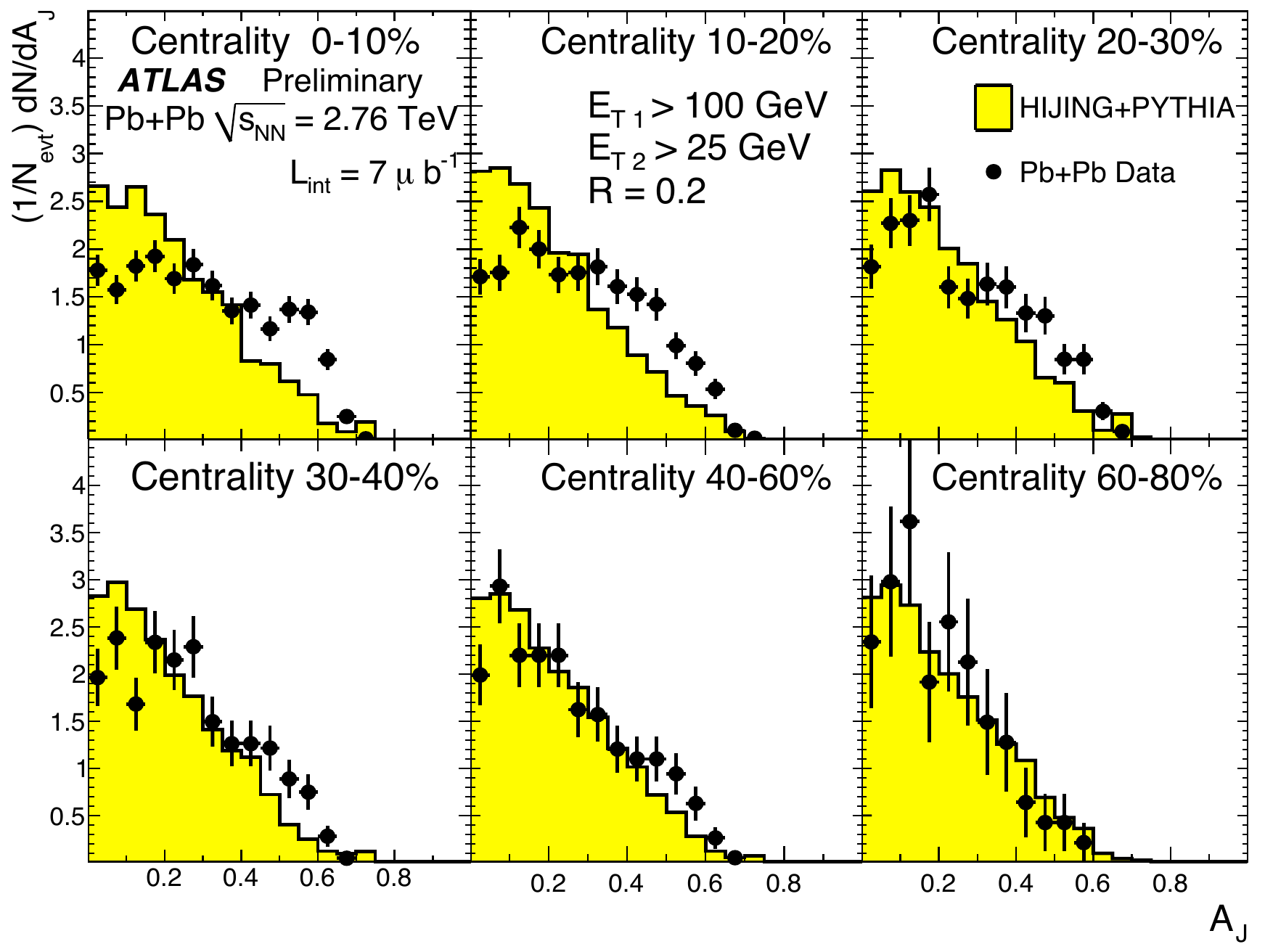}
\vspace{-8mm}
\end{center}
     \caption{Dijet transverse energy balance characterized by the asymmetry parameter $A_j$=\((E_{T1}-E_{T2}) / (E_{T1}+E_{T2})\) where \(E_{Ti}\) is the transverse energy of jet~i, measured by ATLAS in \pbpb collisions at \sqn = 2.76 TeV for various centrality classes \cite{jets-atlas}. }
 \label{fig:jets-momentum-balance-atlas}
\vspace{-3mm}
  \end{figure}

$\bullet$  \underline {Energy and momentum balance.}
At LHC, dijets exhibit a large transverse energy imbalance that grows with centrality as first shown by ATLAS (see Fig. \ref{fig:jets-momentum-balance-atlas}) \cite{jets-atlas}. Very similar results are also reported by CMS \cite{jets-energy-balance-cms}. 
 Detailed studies carried out by CMS demonstrate that  this imbalance  results from the incomplete accounting of the jet energy by the jet algorithm. It is a direct consequence of the soft particles produced as a result of the energy lost by high energy partons as they traverse the medium, that are not included in the jet algorithm because they fall outside the jet cone or below the energy threshold. 

\begin{figure}[h!]
      \includegraphics[width=70mm]{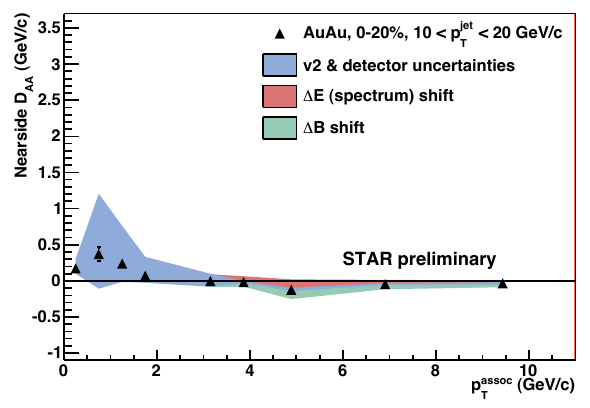}
     \includegraphics[width=70mm]{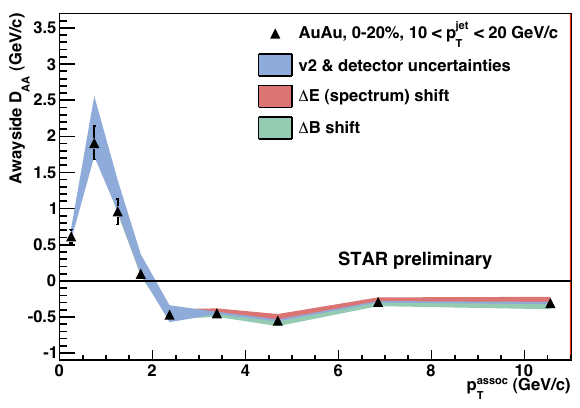}
    \vspace{-3mm}
   \caption{Near-side (left) and away-side (right) $D_{AA}$ distributions as a function of the associated hadron transverse momentum \(p_T^{assoc}\) in central Au+Au collisions measured by STAR in central \auau collisions at \sqnr~\cite{caines-qm11}. See text for the definition of $D_{AA}$.}
                     \label{fig:jets-momentum-difference-star}
     \vspace{-4mm}
 \end{figure}

Energy and momentum balance studies are also carried out at RHIC.   
Fig. \ref{fig:jets-momentum-difference-star} shows the transverse momentum difference between AA and pp collisions, \( D_{AA}(p_T^{assoc}) = [Y_{AA}(p_T^{assoc}) - Y_{pp}(p_T^{assoc})] p_T^{assoc} \) (where \(Y_{aa}(p_T^{assoc})\) represents the yield of associated tracks with transverse momentum \(p_T^{assoc}\) in aa collisions)  for the near-side (left panel) and the away-side (right panel) jets \cite{caines-qm11}.
For the near-side jet, $D_{AA}$ is consistent with zero, i.e. the momentum profile is very similar in central Au+Au and pp collisions, there are no changes as expected from the surface bias effect of the jet trigger. On the other hand, for the away side jet there is a deficit of high 
\pt~tracks that is within errors compensated by an excess of low \pt~tracks.  

\begin{figure}[h!]
     \begin{center}
           \includegraphics[width=120mm]{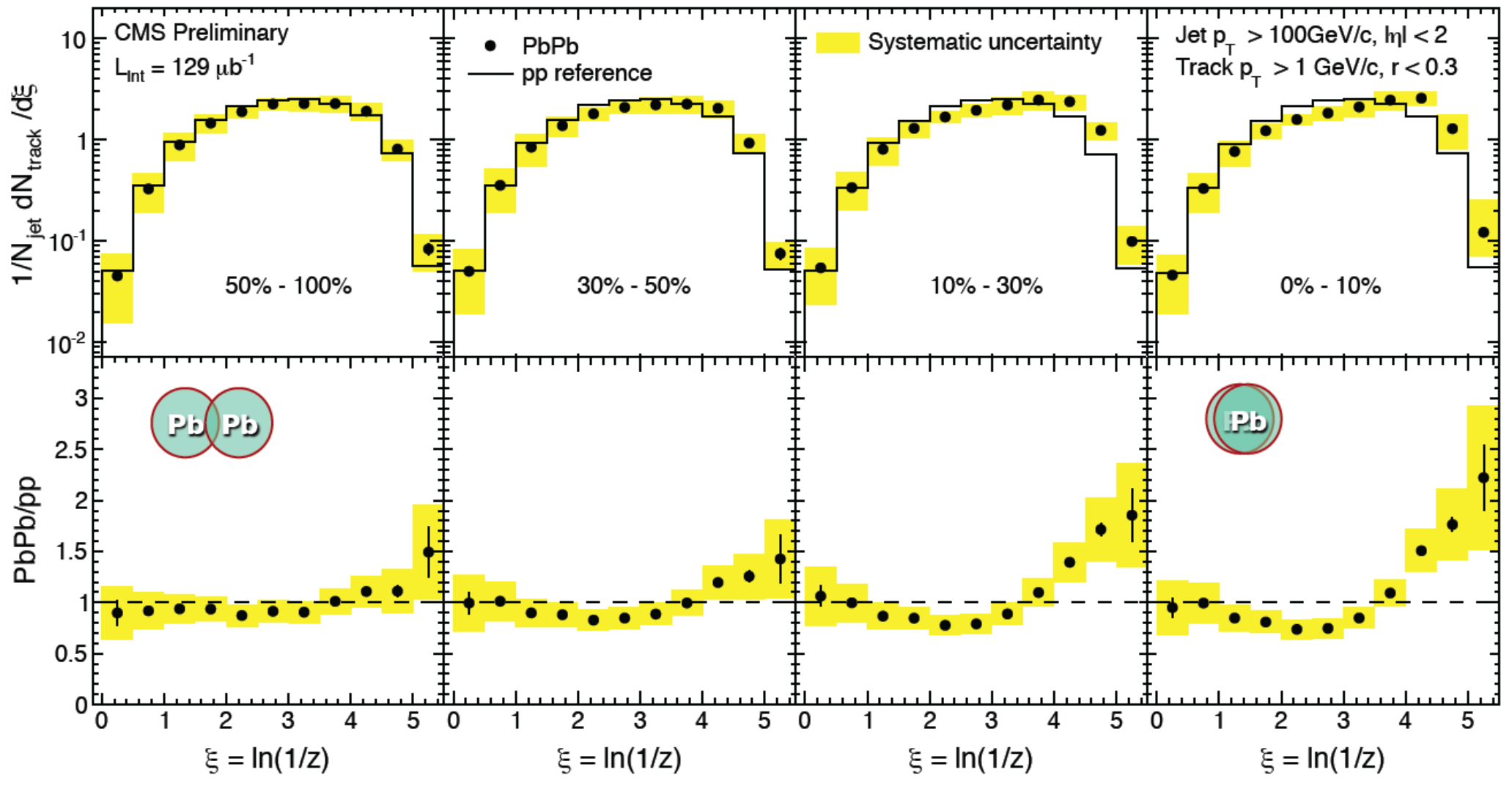}
     \end{center}
\vspace{-6mm}
     \caption{Fragmentation functions in pp and for several centrality classes in Pb+Pb collisions measured at LHC by CMS using charged particle tracks inside the jet cones. $z$ represents the ratio of the track momentum along the jet axis to the magnitude of the jet momentum. Inclusive jets with \pt~$>$ 100 GeV/c and tracks with \pt~$>$~1~GeV/c and within a cone of r=0.3 are considered in this analysis \cite{jet-shape-ff-cms}.}
\label{fig:jets-ff-cms}
 \vspace{-6mm}  
\end{figure}

 $\bullet$  \underline {Fragmentation functions.}
The jet properties described above imply a change of the jet fragmentation function in AA collisions compared to pp collisions and indeed 
both ATLAS and CMS have recently reported such changes \cite{jet-ff-atlas, jet-shape-ff-cms}. The top panels of Fig.~\ref{fig:jets-ff-cms} show the fragmentation functions in \pbpb collisions for several centrality bins together with the pp reference fragmentation function \cite{jet-shape-ff-cms}. The bottom panels show the ratio of the \pbpb to pp fragmentation functions. The results are plotted against $\xi = ln(1/z)$ where $z = p^{||}/p_{jet}$ is the ratio of the track momentum along the jet axis to the magnitude of the jet momentum.
For the most peripheral collisions, 50-100\%, the ratio is consistent with unity i.e. there is no modification of the \pbpb  fragmentation function as one would expect.
There is also no modification of the \pbpb fragmentation function at all centralities for  high \pt~tracks with $\xi <$ 1, i.e. when a single track carries more than 40\% of the total jet momentum. For all the other cases there is a clear modification of the fragmentation functions that increases with centrality.  The modification is mainly characterized by an excess  of low \pt~tracks \pt~$<$ 3 GeV/c  ($\xi >$ 3.5), reaching a factor of $\sim$2 for the most central collisions\footnote{This was not seen in the earlier CMS analysis that was restricted to tracks with \pt~$>$ 4 GeV/c and led to the conclusion that the jet fragmentation function is not modified in \pbpb collisions \cite{cms-ff, it-epic}.}. This excess is accompanied by a small deficit of tracks with \pt~$ >$ 3 GeV/c.  

Very similar modifications of the jet fragmentation function were observed in central \auau collisions by PHENIX in gamma-hadron correlations \cite{jets-raa-phenix}.
 
\subsection{Calibrated probes}
\label{calibrated_probes}
The results presented in the previous subsections are a consequence of the energy lost by high energy partons as they traverse the strongly interacting colored medium.  But in all these measurements of dijets, jet-hadron or hadron-hadron correlations, the initial parton conditions are not known. Both partons, produced in the initial hard scattering process, can undergo energy loss and the measurements do not provide  direct information on the amount of energy lost in the medium. 
\begin{figure}[h!]
\begin{center}
\begin{minipage}[b]{0.48\linewidth}
\includegraphics[width=60mm, height=50mm]{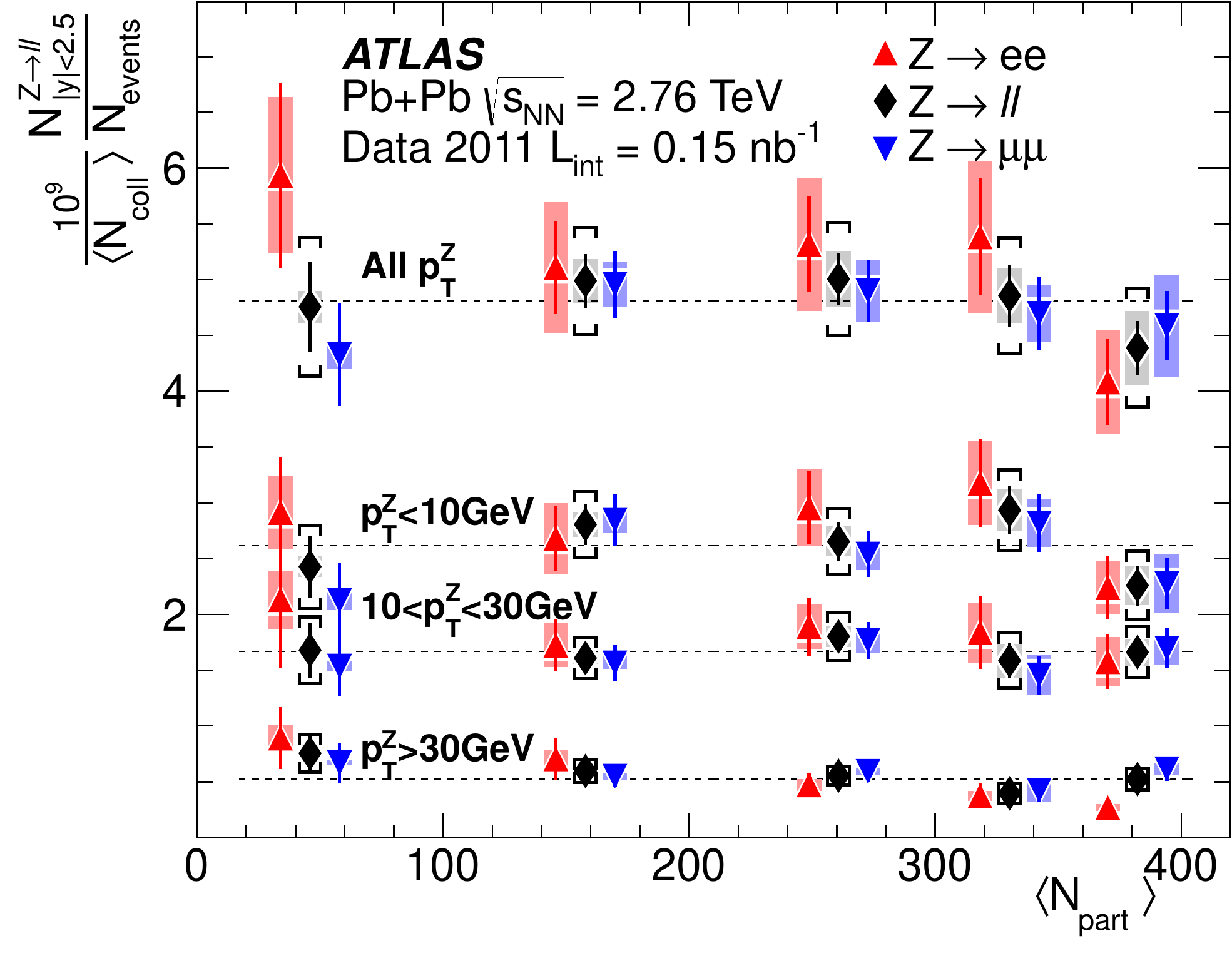}
\vspace{-4mm}
\caption{Z boson yield vs. centrality in \pbpb collisions at \sqn = 2.76 TeV from ATLAS. Results are shown separately for Z$\rightarrow e^+e^-$, Z$\rightarrow \mu^+\mu^-$, the two dilepton channels combined, and several \pt~bins of the Z boson as indicated in the figure \cite{z-atlas}.}
\label{fig:z-atlas}
\end{minipage}
\hspace{0.05\linewidth}
\begin{minipage}[b]{0.45\linewidth}
\includegraphics[width=60mm, height=50mm]{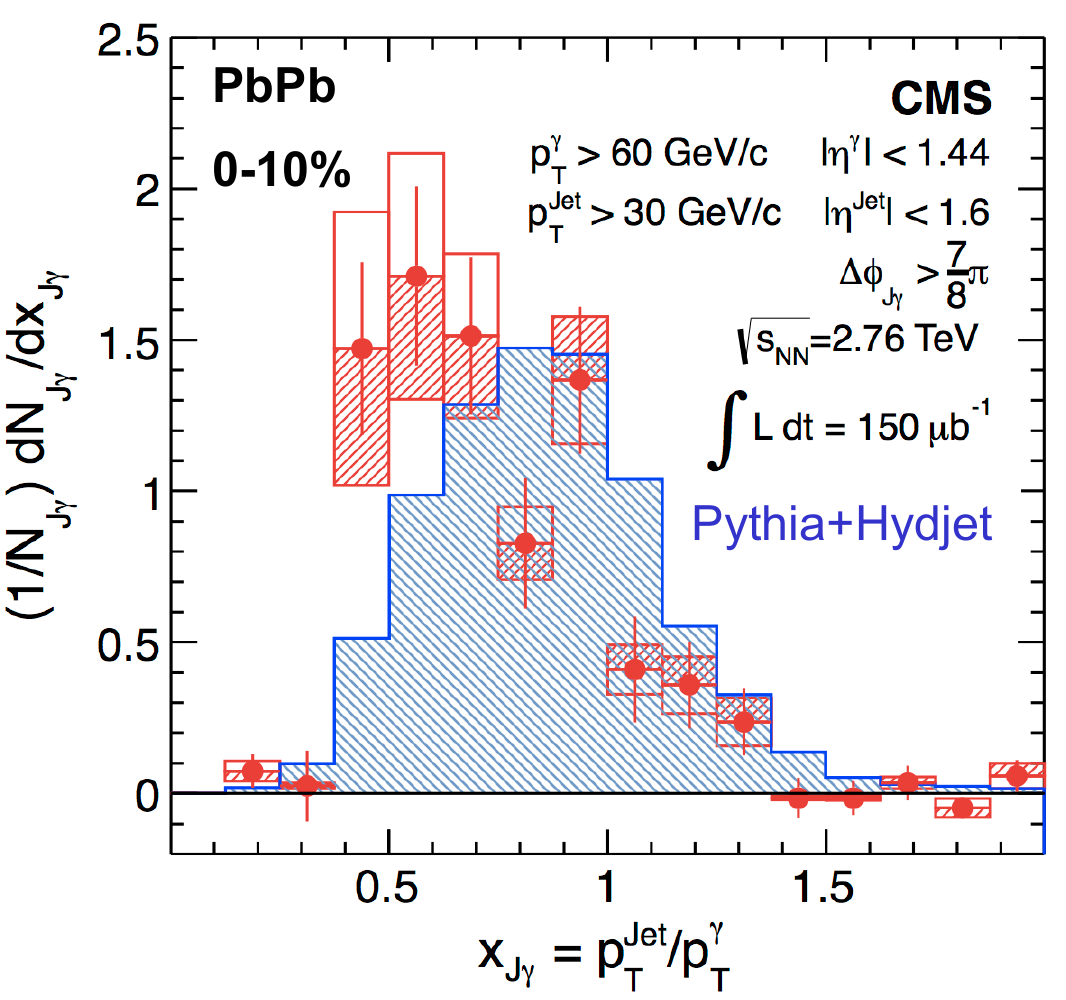}
\vspace{-4mm}
\caption{Fraction of the jet \pt~to $\gamma$ \pt~in $\gamma$-jet studies in central \pbpb collisions at \sqn = 2.76 TeV from CMS. Analysis is restricted to \pt$^\gamma >$ 60 GeV and \pt$^{Jet} >$ 30 GeV. Results are compared to simulations based on Pythia + Hydjet \cite{gamma-jet-cms}.}
\label{fig:gamma-jet-cms}
\end{minipage}
\vspace{-6mm}
\end{center}
\end{figure}
Measurements using calibrated probes, like $\gamma$-jet or Z,W-jet are considered golden channels for a direct measurement of the energy lost in the medium. The $\gamma$ or the  weak bosons Z or W detected through their leptonic decay channels, are not affected by the medium and serve to define the initial energy of the recoiling jet. Also, the surface bias effect inherent to dijet measurements does not exist in these channels. 
But these are not easy measurements. Direct, isolated photons are copiously produced at LHC energies but suffer from large backgrounds originating mainly from $\pi^0$ and other decays. There is also a possible contamination from fragmentation photons. The W bosons are not fully reconstructed. They are identified through single leptons from their $l\nu_l$ decay channel.
Z bosons appear as the best candidate from this point of view as they are practically free of any background and can be fully reconstructed through their dilepton decay channels.  However they suffer from low production cross sections and high luminosity is required for precise Z-jet measurements. 
 
A large experimental effort is ongoing at RHIC and LHC to measure direct photons and at the LHC to measure weak bosons. There is convincing evidence that isolated photons and weak bosons do not interact with the medium.  This is desmonstrated  in the left panel of Fig.~\ref{fig:raa-all-particles} and the right panel of Fig.~\ref{fig:raa-cms} showing that the \raa of photons is unity at RHIC and LHC, respectively.  
 A precise measurement of the Z boson production at the LHC in both the electron and di-muon decay channels shows that the production of Z bosons scales with the number of binary collisions as shown in Fig.\ref{fig:z-atlas}. 
First results on $\gamma$-jets studies at RHIC and LHC were recently reported \cite{jets-raa-phenix, gamma-jet-cms, gamma-jet-atlas}. Fig.~\ref{fig:gamma-jet-cms} shows the fraction x$_{J\gamma}$ of the jet \pt~ to the $\gamma$ \pt~in $\gamma$-jet correlations measured by CMS at LHC in central \pbpb collisions. The analysis is performed for photons with \pt$^\gamma >$ 60 GeV/c and jets with \pt$^{jet} >$ 30 GeV/c. The results show a clear shift of the distribution with respect to simulations based on Pythia + Hidjet \cite{gamma-jet-cms}.

\section{Direct photons}
Electromagnetic probes (real and virtual photons) play a unique role in the diagnosis of the sQGP. They provide information about the plasma temperature and are sensitive to chiral symmetry restoration effects and in-medium properties of hadrons. The most recent RHIC and LHC results on real photons are discussed below. Results on virtual photons are only available at RHIC. For the most recent results see \cite{qm12-proc}.

\begin{figure}[h!]
      \includegraphics[width=70mm, height=55mm]{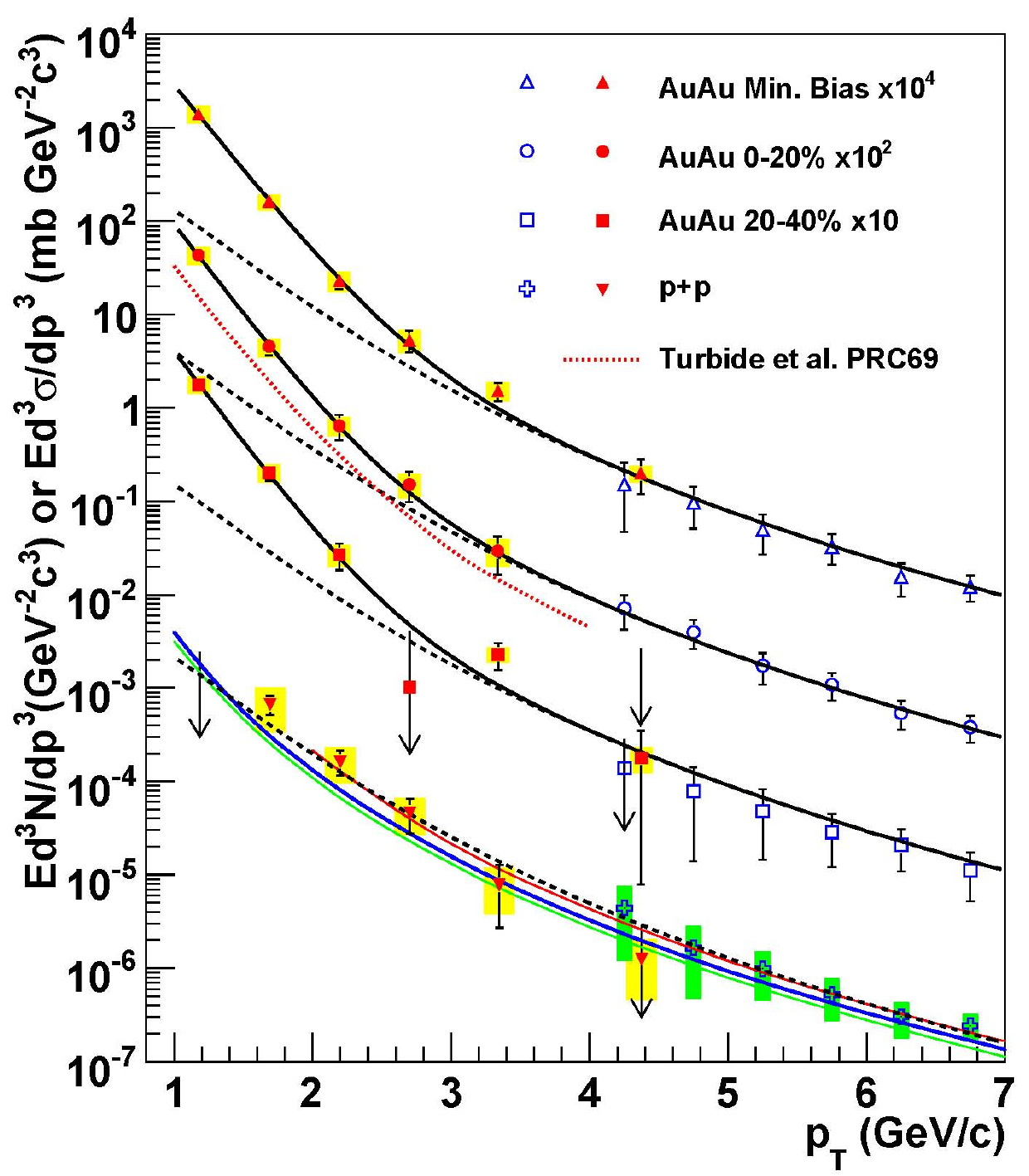}
      \includegraphics[width=70mm, height=55mm]{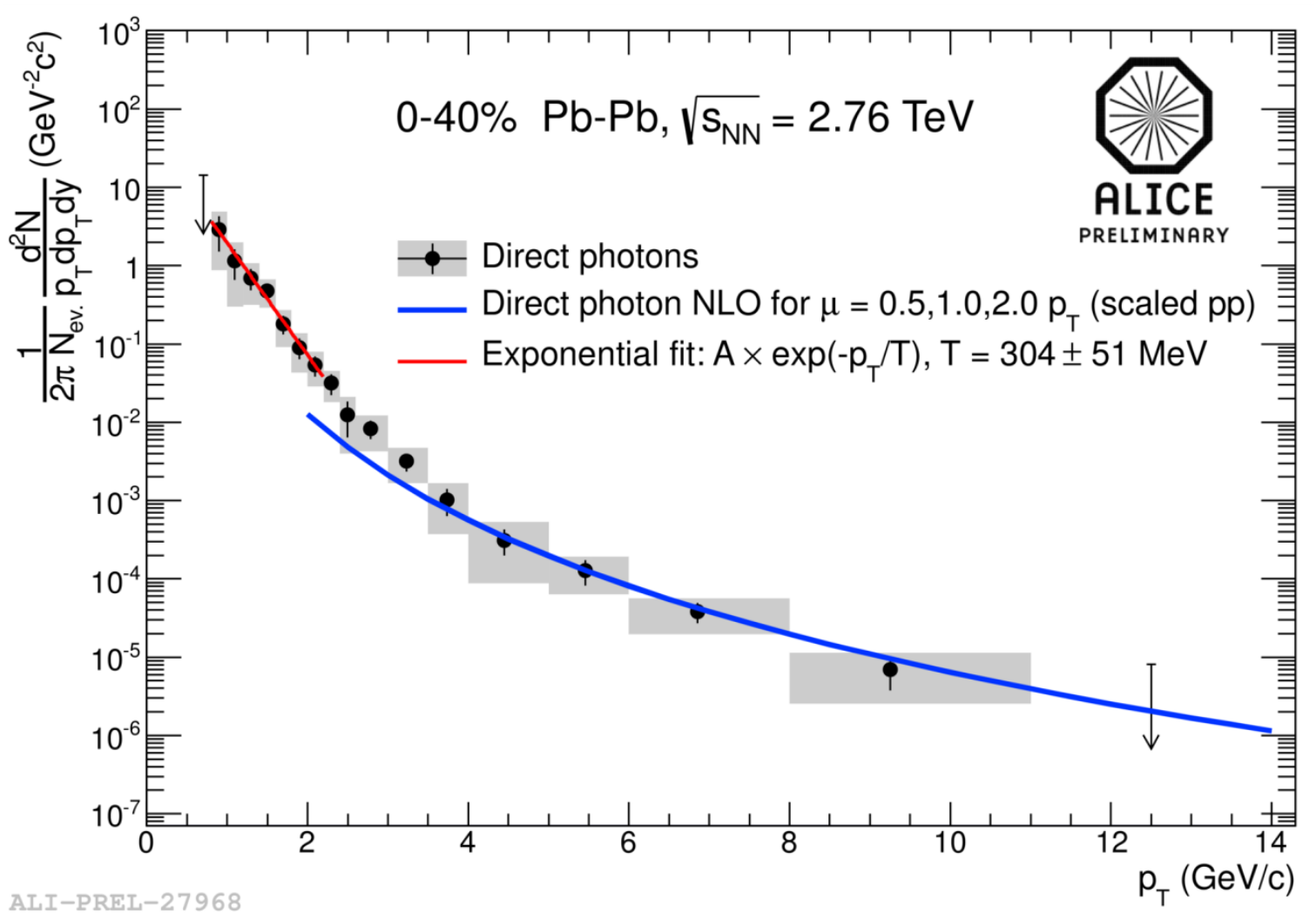}
      \vspace{-3mm}
   \caption{Left panel: Direct photon invariant \pt~spectra in pp and \auau collisions at \sqnr measured by PHENIX. The dashed curve on the pp data represent a fit with a power law function. The black curves on the \auau data are exponential plus the N$_{coll}$ scaled pp fit. For more details see \cite{ppg086-photons-auau}. Right panel: Direct photon invariant \pt~spectrum in \pbpb collisions at \sqn = 2.76 TeV with NLO pQCD calculations at high \pt~and exponential fit at low \pt~\cite{alice-direct-photons}.}
                     \label{fig:direct-photons-phenix}
     \vspace{-3mm}
 \end{figure}

As part of a systematic study of electromagnetic probes at RHIC, the PHENIX experiment has measured direct photons in p+p and \auau collisions at \sqnr~\cite{ppg086-photons-auau}. The results are shown in Fig.~\ref{fig:direct-photons-phenix}. The p+p data are well reproduced by NLO pQCD calculations down to low \pt. On the other hand, the \auau data exhibit a strong photon excess in the \pt~range of 1-3 GeV/c, beyond the  p+p yield scaled by the number of binary collisions N$_{coll}$. This excess has an exponential shape with a slope parameter  T = 221 $\pm 19^{stat} \pm 19^{syst}$ MeV in central (0-20\%) collisions. The excess has been interpreted as thermal radiation from the sQGP  thus providing the first information about the temperature of the medium averaged over the space-time evolution of the collision. Using hydrodynamical models one can infer an initial temperature of T$_{ini}$ = 300 to 600 MeV depending on the assumed formation time ($\tau$ = 0.6-0.15 fm/c) of the system. A similar analysis and a similar excess of direct photons have been recently reported by the ALICE experiment \cite{alice-direct-photons}. The ALICE results are displayed in the right panel of Fig.~\ref{fig:direct-photons-phenix}. The slope parameter of the excess is T = 305 $\pm$ 51 $^{stat+syst}$ MeV,  higher than at RHIC as one would expect if the excess is indeed the thermal radiation from the sQGP.

In order to further strengthen the interpretation in terms of thermal radiation, PHENIX measured the direct photons in \dau collisions \cite{ppg140-photons-dau}. 
 The direct photon spectrum in \dau collisions is well reproduced by a fit to the pp data scaled with the number of binary collisions and there is no excess of low \pt~photons. 
This result clearly demonstrates that the large excess of direct photons observed in \auau at low \pt is not due to initial state effects and reinforces the interpretation of the \auau excess as thermal radiation from the sQGP.

\begin{figure}[h!]
  \begin{center}
     \includegraphics[width=70mm]{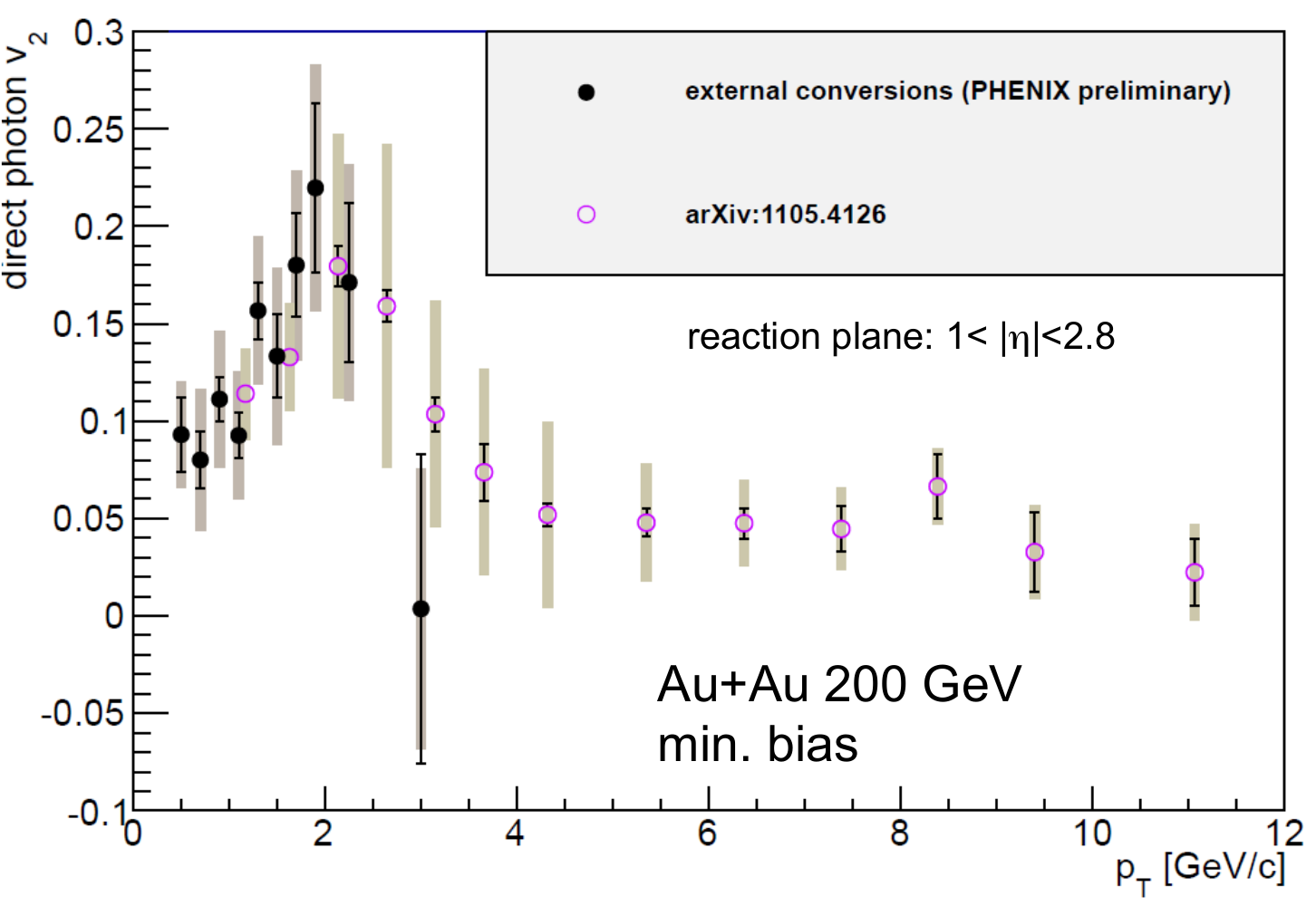}
   \end{center}
    \vspace{-6mm}
\caption{ Direct photons \vtwo~results obtained by PHENIX in minimum bias \auau collisions at \sqnr~using two completely independent analysis methods based on inclusive photons and on external conversion photons, respectively \cite{it-qm12}.}
                     \label{fig:phenix-v2-photons}
     \vspace{-6mm}
 \end{figure}

On the other hand PHENIX has also measured a large elliptic flow  \vtwo~of low \pt~direct photons in \auau collisions at \sqnr, comparable in shape and magnitude to the elliptic flow of pions \cite{phenix-photons-v2}. The large \vtwo~of direct photons has been confirmed in a completely  independent  analysis based on external conversion photons \cite{it-qm12}. The new analysis allowed to extend the \pt~range of the measurement down to 0.5 GeV/c. The results are shown in Fig.~\ref{fig:phenix-v2-photons} \cite{it-qm12}. The large \vtwo~values of direct photons is a surprising and challenging result. The direct photon yield at low \pt~
is dominated by the excess mentioned above that is interpreted as thermal radiation from the sQGP. However thermal radiation implies early emission whereas a large \vtwo~implies late emission. Assuming that all measurements are correct, these two seemingly contradicting features call for a revision of the current interpretation of the results.

\section{Summary and Outlook}
The field of relativistic heavy ion physics is now in a phase of precision measurements to characterize the properties of the sQGP. Only a few topics could be discussed in the limited scope of this paper, covering the most recent results on flow, energy loss and direct photons. The field benefits from  unique opportunities offered by five large experiments operating at two large facilities and producing high quality  and complementary  data. Results are obtained at unprecedented rate and considerable progress should be achieved in the next few years. 
 
{\centering \section*{ACKNOWLEDGMENTS}}
The author acknowledges the conference organizers for the opportunity to present this talk and for a most stimulating conference. This work was supported by  the Israeli Science Foundation and the Nella and Leon Benoziyo Center for High Energy Physics. 

%
%

\end{document}